\documentclass[journal=jceaax,manuscript=article]{achemso}
\usepackage{chemformula} 
\usepackage{}
\usepackage{bm}
\usepackage[T1]{fontenc} 
\usepackage{subcaption}
\usepackage{amssymb}
\usepackage{multirow}
\usepackage{longtable}
\usepackage{rotating}
\usepackage{hhline}
\usepackage{amsmath}
\usepackage{gensymb}
\usepackage{graphicx}
\usepackage{color}
\usepackage{tikz}
\usetikzlibrary{shapes.geometric}
\usepackage{booktabs}
\usepackage{siunitx}
\usepackage{lineno}
\author{Nicolas~Hayer}
\affiliation{Laboratory of Engineering Thermodynamics, RPTU Kaiserslautern, Erwin-Schrödinger-Str. 44, 67663 Kaiserslautern, Germany}

\author{Thomas~Specht}
\affiliation{Laboratory of Engineering Thermodynamics, RPTU Kaiserslautern, Erwin-Schrödinger-Str. 44, 67663 Kaiserslautern, Germany}

\author{Justus~Arweiler}
\affiliation{Laboratory of Engineering Thermodynamics, RPTU Kaiserslautern, Erwin-Schrödinger-Str. 44, 67663 Kaiserslautern, Germany}

\author{Dominik~Gond}
\affiliation{Laboratory of Engineering Thermodynamics, RPTU Kaiserslautern, Erwin-Schrödinger-Str. 44, 67663 Kaiserslautern, Germany}

\author{Hans~Hasse}
\affiliation{Laboratory of Engineering Thermodynamics, RPTU Kaiserslautern, Erwin-Schrödinger-Str. 44, 67663 Kaiserslautern, Germany}

\author{Fabian~Jirasek\textsuperscript}
\affiliation{Laboratory of Engineering Thermodynamics, RPTU Kaiserslautern, Erwin-Schrödinger-Str. 44, 67663 Kaiserslautern, Germany}
\email{fabian.jirasek@rptu.de}

\title{Prediction of Activity Coefficients by Similarity-Based Imputation using Quantum-Chemical Descriptors}

\begin{document}

\begin{abstract}
In this work, we introduce a novel approach for predicting thermodynamic properties of binary mixtures, which we call the similarity-based method (SBM). The method is based on quantifying the pairwise similarity of components, which we achieve by comparing quantum-chemical descriptors of the components, namely $\sigma$-profiles. The basic idea behind the approach is that mixtures with similar pairs of components will have similar thermodynamic properties. The SBM is trained on a matrix that contains some data for a given property for different binary mixtures; the missing entries are then predicted by the SBM. As an example, we consider the prediction of isothermal activity coefficients at infinite dilution ($\gamma^\infty_{ij}$) and show that the SBM outperforms the well-established physical methods modified UNIFAC (Dortmund) and COSMO-SAC-dsp. In this case, the matrix is only sparsely occupied, and it is shown that the SBM works also if only a limited number of data for similar mixtures is available. The SBM idea can be transferred to any mixture property and is a powerful tool for generating essential data for many applications.
\end{abstract}

\section{Introduction}
Thermodynamic properties of mixtures are fundamental for the design and optimization of processes. In this work, we describe a novel approach for predicting properties of binary mixtures based on \textit{similarities} between components. This novel similarity-based method (SBM) is built on the fundamental assumption that similar components exhibit similar properties ({\it similia similibus solvuntur}), making component similarities highly informative inputs for predictive thermodynamic models. 

Molecular similarity is commonly used in computational chemistry and pharmaceutical research for database searching and component selection in high-throughput screening. The goal of these applications is to find components that exhibit a behavior that is similar to that of a reference component with desired properties. This is achieved by identifying similar substructures or calculating overall similarity measures, resulting in a list of the most similar molecules in the database and, ultimately, guiding drug discovery and optimization. To perform these pairwise molecular comparisons, a molecular representation of the components and a method to evaluate the similarity based on these representations are required. Various approaches have been proposed for this purpose in the literature, each with its own merits and limitations~\cite{Nikolova.2003, Stumpfe.2011}.

The most common molecular representations for similarity searches are molecular fingerprints, which encode structural information into bit vectors, such as the presence of specific functional groups\cite{Flower.1998,Stumpfe.2011}. Analyzing fingerprint similarities is computationally efficient, as it only involves comparing bit strings. The Tanimoto coefficient is the most popular metric for assessing fingerprint similarity~\cite{Flower.1998, Fligner.2002, Bajusz.2015}. Other molecular representations for assessing similarity include molecular graphs, molecular descriptor vectors, SMILES, SMARTS, and pharmacophores~\cite{Raymond.2002, Nikolova.2003, Stumpfe.2011}. Molecular descriptors based on quantum-chemical charge distribution calculations, such as $\sigma$-profiles~\cite{Klamt.1995}, are rarely used to assess similarities in pharmaceutical research, despite their potential~\cite{Thormann.2012, Thormann.2024}.

While the idea of using similarities is implicitly at the heart of many models for predicting thermodynamic properties for unstudied systems, our similarity-based method (SBM) exploits that idea based on a measure of similarity directly. 

Among the thermodynamic properties of mixtures, the activity coefficient is particularly significant since it quantifies the non-ideality of liquid mixtures, which is essential for accurately modeling reaction and phase equilibria~\cite{Brouwer.2019}. A highly informative limiting case is the activity coefficient $\gamma^\infty_{ij}$ of a solute $i$ infinitely diluted in a solvent $j$, as many mixture properties can be predicted based on the knowledge of the limiting activity coefficients. However, despite their importance, experimental data for $\gamma^\infty_{ij}$ are scarce, even in comprehensive databases for thermophysical properties such as the Dortmund Data Bank~\cite{DDB2023}, due to the high cost and time required for their measurement~\cite{Orbey.1991, Kojima.1997}. Consequently, reliable prediction methods are essential.

Activity coefficients are usually calculated from models of the Gibbs excess energy $G^\mathrm{E}$. Predictions for binary mixtures, for which no data are available, can be obtained from group contribution methods, namely UNIFAC~\cite{Fredenslund.1975, Wittig.2003} and modified UNIFAC (Dortmund)~\cite{Weidlich.1987, Constantinescu.2016}, or using the COSMO-RS approach~\cite{Klamt.1995, Klamt.2000, Klamt.2005}, which is based on quantum-chemical component descriptors, the $\sigma$-profiles. Open-source versions of COSMO-RS include COSMO-SAC~\cite{Lin.2002, Hsieh.2010} and COSMO-SAC-dsp~\cite{Hsieh.2014}. The $\sigma$-profiles describe the screening charge density of a molecule embedded in an electrically conductive continuum by a probabilistic distribution $p(\sigma)$ across the molecule’s surface segments, where $\sigma$ is the charge of the segment~\cite{Klamt.1995}.

In addition to these physical prediction methods, new machine learning (ML) methods and hybrid models that combine physics with ML have been developed recently~\cite{Jirasek.2021,Jirasek.2023b}. These methods include graph neural networks (GNN)~\cite{SanchezMedina.2022}, transformer models~\cite{Winter.2022}, and matrix completion methods (MCM)~\cite{Jirasek.2020,Jirasek.2020b,Damay.2021}. Additionally, many ML methods have been developed to predict activity coefficients over the entire concentration range, which could also be applied to the special case of activity coefficients at infinite dilution.~\cite{Winter.2023,Jirasek.2023,Rittig.2023,Specht.2024,Hayer.2024}.

We apply the SBM here to predict activity coefficients at infinite dilution $\gamma^\infty_{ij}$ in binary mixtures. The SBM thereby relies on two sources of information: a novel similarity measure $S_{mn}$ between two components $m$ and $n$ and available experimental data for $\gamma^\infty_{ij}$. The similarity measure $S_{mn}$ is based on a comparison of $\sigma$-profiles of the pair of components and used to screen the experimental database, identifying $\gamma^\infty_{ij}$ values from similar mixtures that are then used for predictions by imputation. We benchmark the developed SBM with modified~UNIFAC (Dortmund)~\cite{Constantinescu.2016}, COSMO-SAC~\cite{Hsieh.2010}, and COSMO-SAC-dsp~\cite{Hsieh.2014} as three well-established physics-based methods for predicting $\gamma^\infty_{ij}$. We emphasize that the SBM for predicting $\gamma^\infty_{ij}$ is an example; the approach is generic and can be transferred to any other binary property.

\clearpage
\section{Database}
Experimental data on activity coefficients at infinite dilution in binary mixtures, $\gamma^\infty_{ij}$, were obtained from the Dortmund Data Bank (DDB)~\cite{DDB2023}. In the preprocessing step, all data sets containing undefined components or labeled as "poor quality" by the DDB were discarded. The focus was restricted to binary mixtures at a temperature of $T = 298.15 \pm 1$ K. If multiple measurements existed for the same binary mixture, the median of these values was adopted. For scaling purposes, the logarithmic activity coefficients, $\ln\gamma^\infty_{ij}$, were used throughout this study.

The proposed SBM uses $\sigma$-profiles obtained from quantum-chemical COSMO calculations to calculate the similarity between two components. In this work, the $\sigma$-profiles were taken from the open-source database provided by Bell et al.~\cite{Bell.2020}, which features results for 2,261 different components. Components not available in this database were excluded from our data set.

\begin{figure}[H]
	\centering
	\includegraphics[width=0.5\textwidth]{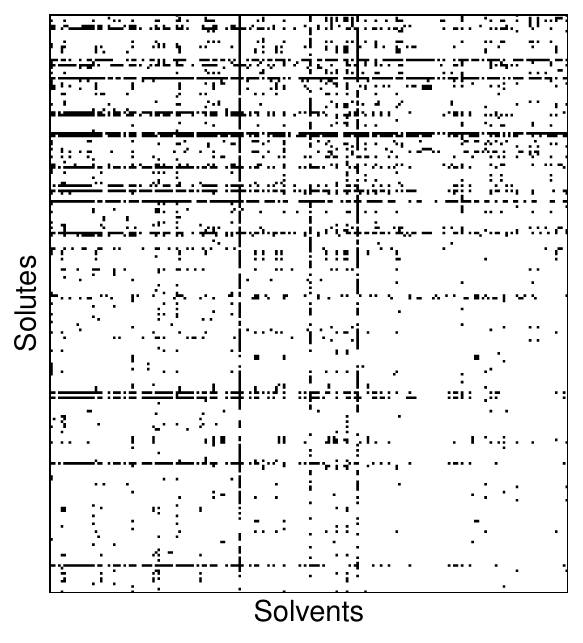}
	\caption{Matrix representing the experimental data on logarithmic activity coefficients at infinite dilution $\ln \gamma^\infty_{ij}$ for binary mixtures at 298.15$\pm$1~K from the DDB~\cite{DDB2023} after preprocessing (see text). Experimental data are available for 3,568 binary mixtures, constituting about 8\% of all possible combinations of the considered 221 solutes and 198 solvents.}
	\label{Sim1_fig:Heatmap_Exp}
\end{figure}

Finally, for evaluating the model using leave-one-out analysis, at least two experimental data points were required for each solute and solvent; therefore, data for which this condition was violated were removed. The final data set is visualized in Fig.~\ref{Sim1_fig:Heatmap_Exp} and comprises 3,568 data points for $\gamma^\infty_{ij}$, covering 221 solutes and 198 solvents.

\clearpage
\section{Similarity-Based Method}
\subsection{Similarity Score}
\begin{figure}[H]
    \centering
    \includegraphics[width=\textwidth]{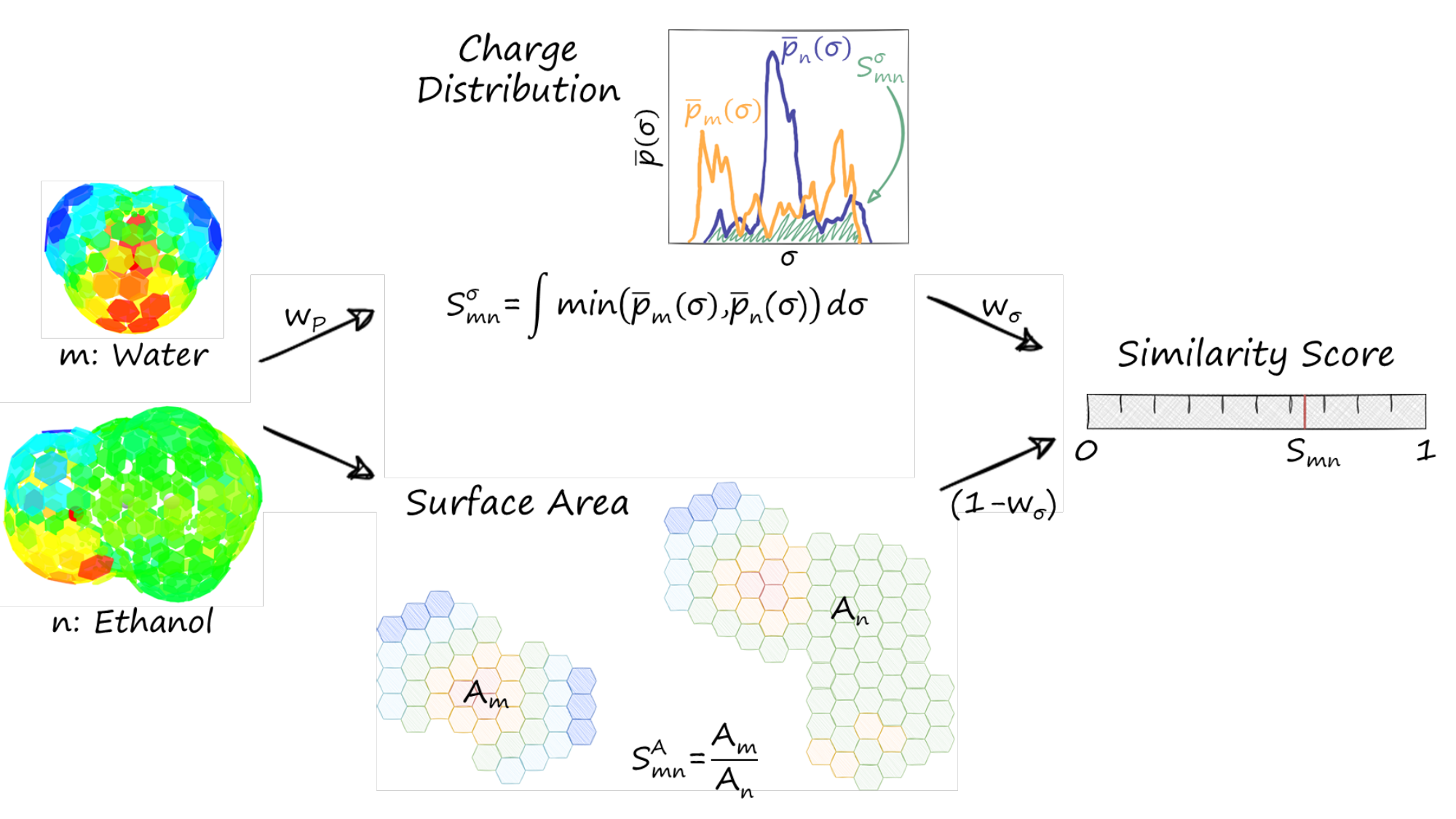}
    \caption{Schematic depiction of calculating the similarity between two components (water and ethanol in this example) as proposed in this work. The final similarity score $S_{mn}$ is composed of two contributions: a similarity based on charge distribution $S^\sigma_{mn}$ and a size similarity derived from the surface areas $S^A_{mn}$, which are combined in a weighted sum.}
    \label{Sim1_fig:Scheme}
\end{figure}

Here, we introduce a novel similarity score $S_{mn}$ between two components $m$ and $n$ based on quantum-chemical COSMO calculations. The score $S_{mn}$ is scaled such that its values range from 0 (highly dissimilar components) to 1 (highly similar components) and consists of two contributions, as also indicated in Fig.~\ref{Sim1_fig:Scheme}: the similarity based on surface charge distributions $S^\sigma_{mn}$ and the similarity of the surface area $S^A_{mn}$ as it is also used in the COSMO method; $S^\sigma_{mn}$ and $S^A_{mn}$, which are described in detail in the following, are also defined to range from 0 to 1. The final similarity score $S_{mn}$ is obtained from a weighted sum of $S^\sigma_{mn}$ and $S^A_{mn}$:
\begin{equation}
\label{Sim1_eq:SimScore}
S_{mn} = w_\sigma \cdot S^\sigma_{mn} + (1-w_\sigma) \cdot S^A_{mn}
\end{equation}
where $w_\sigma$ is the weighting factor that controls the relative importance of the surface charge distribution similarity compared to the surface area similarity.

The size similarity $S^A_{mn}$ is defined as the cavity surface area $A$ of the smaller molecule divided by the one of the larger molecule:
\begin{align}
\label{Sim1_eq:SimScore_A}
S^A_{mn} &= \begin{cases} 
\frac{A_{m}}{A_{n}}, & \text{if } A_{m} < A_{n} \\
\frac{A_{n}}{A_{m}}, & \text{if } A_{m} > A_{n}
\end{cases}
\end{align}

For the similarity of the surface charge distributions $S^\sigma_{mn}$, the overlapping proportion of the $\sigma$-profiles of the two components is used, which is calculated using discrete bins for $\sigma$ via:
\begin{equation}
S^\sigma_{mn} = \sum\limits_{k=1}^{N_\sigma} \mathrm{min}\left(\bar{p}_m(\sigma_k),\bar{p}_n(\sigma_k)\right)
\label{Sim1_eq:SimScore_sigma}
\end{equation}
where $\bar{p}_m(\sigma_k)$ and $\bar{p}_n(\sigma_k)$ are modified $\sigma$-profiles, preprocessed as described in the following. All $\sigma$-profiles are given here in a discretized version with $\sigma$ being divided into $N_\sigma=51$ bins ranging from -0.025~e$\si{\angstrom}^{-2}$ to 0.025~e$\si{\angstrom}^{-2}$ with a constant step size of 0.001~e$\si{\angstrom}^{-2}$. We will refer to these values as $\sigma_k$ for $k = 1, \ldots, 51$. Thus, $p_m(\sigma_k)$ is the fraction of the surface area of the component $m$ associated with the screening charge density $\sigma_k$.

We modify the $\sigma$-profiles by introducing $w_\mathrm{P}$, which is applied to control the weight on the polar regions in the $\sigma$-profiles by being either 0 (no influence) or 2 (more focus on polar regions):
\begin{equation}
    p_m^*(\sigma_k) =  p_m(\sigma_k) \cdot (10^3\sigma_k)^{w_\mathrm{P}}
    \label{Sim1_eq:sigma_profiles_weighted}
\end{equation}
By setting $w_\mathrm{P}=2$, the similarity calculation emphasizes charge-dense regions, which can be crucial in cases where the behavior of the components is mainly determined by polar interactions.

In the case of $w_\mathrm{P}=2$, the resulting $p_m^*(\sigma_k)$ does not integrate to 1. Therefore, it is again normalized:
\begin{equation}
    p_m^{**}(\sigma_k) = \frac{p_m^*(\sigma_k)}{\sum\limits_{k=1}^{N_\sigma} p_m^*(\sigma_k)}
    \label{Sim1_eq:sigma_profiles_normalized}
\end{equation}

In the final processing step, we address a potential issue associated with discretized $\sigma$-profiles. Specifically, when calculating the similarity score by comparing the $\sigma$-profiles of two molecules bin-wise, small shifts in $\sigma$ can prevent the detection of structurally similar molecules. Therefore, a moving average with a sliding window of width 2 (corresponding to 0.002~e$\si{\angstrom}^{-2}$) is applied to all profiles to increase the robustness:
\begin{equation}
\bar{p}_m(\sigma_k) = \frac{p_m^{**}(\sigma_{k-1}) + p_m^{**}(\sigma_k)}{2}
\label{Sim1_eq:moving_average}
\end{equation}
The resulting $\sigma$-profiles $\bar{p}_m(\sigma_k)$ are used for calculating the similarity of the surface charge distributions $S^\sigma_{mn}$ (see Eq.~(\ref{Sim1_eq:SimScore_sigma})). Together with the similarity of the surface area $S^A_{mn}$ (see Eq.~(\ref{Sim1_eq:SimScore_A})), the final similarity score $S_{mn}$ is calculated (see Eq.~(\ref{Sim1_eq:SimScore})).

The two introduced weights $w_\sigma$ (in Eq.~(\ref{Sim1_eq:SimScore})), and $w_\mathrm{P}$ (in Eq.~(\ref{Sim1_eq:sigma_profiles_weighted})) are hyperparameters, which were determined by a grid search. The value ranges of the hyperparameters explored in the grid search are detailed in the "Studied Model Variants" section. In addition to these two weights, other modifications to the calculation of $S^\sigma_{mn}$ (e.g.,~emphasizing hydrogen-bonding surface segments) and of $S^A_{mn}$ (e.g.,~including component volume) were tested in preliminary studies, but showed no significant impact on the performance of the SBM and were, therefore, discarded.

\subsection{Prediction of Activity Coefficients}
In this section, we explain how the similarity score defined in the previous section is applied for predicting activity coefficients at infinite dilution $\ln\gamma^{\infty}_{ij}$ in unstudied mixtures, where, basically, the $\ln\gamma^{\infty}_{ij}$ is just an example for a property of a binary mixture. The respective method introduced is called the similarity-based method (SBM). The central idea of the SBM is to find mixtures similar to the unstudied mixture that is of interest but for which experimental data on $\ln\gamma^{\infty}_{ij}$ are available. The activity coefficient in the unstudied mixture, $\ln\gamma^{\infty,\mathrm{pred}}_{ij}$, is then predicted simply by arithmetically averaging the corresponding experimental values $\ln\gamma^{\infty,\mathrm{exp}}_{ij}$ of all similar mixtures.

Here, a \textit{similar mixture} is defined as one with the same solute $i$ (or the same solvent $j$) but a different solvent $n$ (a different solute $m$) for which the similarity score $S_{nj}$ ($S_{mi}$) is higher than a predefined threshold $\xi$, i.e., $S_{nj}>\xi$ ($S_{mi}>\xi$). Consequently, at least one similar mixture for which an experimental data point is available must be in the training set to make a prediction. As a result, there will always be a trade-off when applying the SBM: increasing the threshold value $\xi$ will increase the accuracy, but it will lower the range of applicability. Vice versa, decreasing the value of $\xi$ will increase the range of applicability but decrease the accuracy. 

A leave-one-out approach\cite{Hastie.2017} was applied to assess the SBM to guarantee true predictions. These predictions are also used in comparing the SBM results with the physical benchmark models, which results in a bias in favor of the physical models, as they were very likely also trained with at least some of the data considered here. All calculations of the present study were carried out using Matlab~\cite{MATLAB}.

\subsection{Studied Model Variants}
The SBM described in the previous sections uses two weights, $w_\sigma$ and $w_\mathrm{P}$, in calculating the similarity score $S_{mn}$. These weights were varied in a grid search to explore their effects on model performance. Specifically, $w_\sigma$ was varied from 0 to 1 in increments of 0.1, while $w_\mathrm{P}$ was set to either 0 or 2. This setup resulted in 22 distinct SBM configurations, each representing a different approach to the $S_{mn}$ calculation. The goal of this grid search was to identify the SBM (i.e., weight combination) that performs best for two, often conflicting, objectives: optimizing the accuracy in predicting $\ln\gamma^\infty_{ij}$ in terms of mean absolute error (MAE) and maximizing the scope, i.e., the number of predictable mixtures.

The best-performing SBM, according to these objectives, retains one further adjustable hyperparameter: the threshold $\xi$, which allows users to balance the trade-off between accuracy and scope. Increasing $\xi$ typically results in more accurate predictions but limits the number of predictable data points since higher similarities are demanded for making predictions. Conversely, lowering $\xi$ increases the number of predictable points but reduces the predictive accuracy since data for less similar components are used for the predictions. To assess the impact of $\xi$, it was varied from 0.5 to 1 in increments of 0.01 for each of the 22 SBM configurations.

\clearpage
\section{Results and Discussion}
\subsection{Overall Performance of Different Similarity-Based Methods}
\begin{figure}[H]
    \centering
    \includegraphics[width=0.5\textwidth]{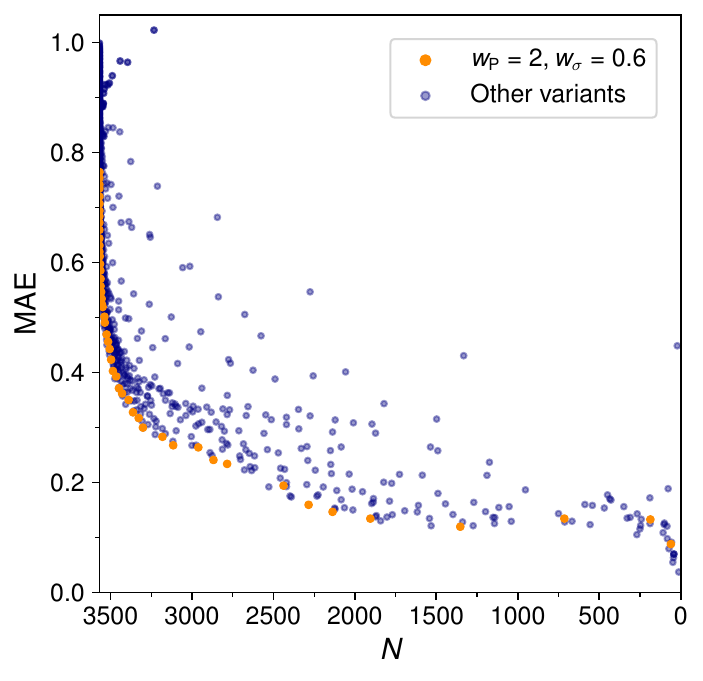}
    \caption{Mean absolute error (MAE) of the predicted $\ln \gamma^\infty_{ij}$ from the leave-one-out analysis over the number of predictable experimental data points $N$ for all tested SBM variants. The results of the best-performing SBM (as specified with the weights $w$) are highlighted in orange.}
    \label{Sim1_fig:GridSearch_MAE}
\end{figure}

Fig.~\ref{Sim1_fig:GridSearch_MAE} shows the predictive accuracy in terms of the MAE of the predicted $\ln\gamma^\infty_{ij}$ over the number of predictable data points $N$ from our data set for all tested SBM variants (by varying the weights and $\xi$).
The model variants in Fig.~\ref{Sim1_fig:GridSearch_MAE} scatter across a broad range of MAE and $N$, underscoring the substantial impact of the selected hyperparameters on model performance. This range highlights the inherent trade-off between predictive accuracy and scope, representing a Pareto optimization problem. In such cases, a solution is considered Pareto-optimal if no feasible solution improves at least one objective without worsening another. Here, certain hyperparameter combinations yield Pareto-optimal SBM variants that achieve maximum accuracy for a given scope and vice versa. The set representing all Pareto-optimal solutions is called the Pareto front.

One particular SBM (with variable $\xi$) consistently lies on or near the Pareto front, highlighted in orange in Fig.~\ref{Sim1_fig:GridSearch_MAE}.
This "best" SBM, defined by $w_\sigma = 0.6$ and $w_\mathrm{P} = 2$, requires only the final tuning of $\xi$ by users to achieve a near-optimal solution tailored to their specific preferences.

The balanced value of $w_\sigma = 0.6$ in the best SBM indicates that components must share similarities in both surface charge distribution and surface area to exhibit comparable values of $\ln\gamma^\infty_{ij}$. Furthermore, $w_\mathrm{P} = 2$ emphasizes the importance of a similar surface charge distribution in the polar regions of the components for similar $\ln\gamma^\infty_{ij}$.

Figs.~S.1 and S.2 in the Supporting Information show further analysis of specific hyperparameter choices. The similarity scores calculated by the best SBM can be used to identify the most similar components for a target component, as exemplified in the Supporting Information (cf.~Tables S.2 and S.3).

\subsection{Comparison to Physical Benchmark Models}
\begin{figure}[H]
    \centering
    \includegraphics[width=\textwidth]{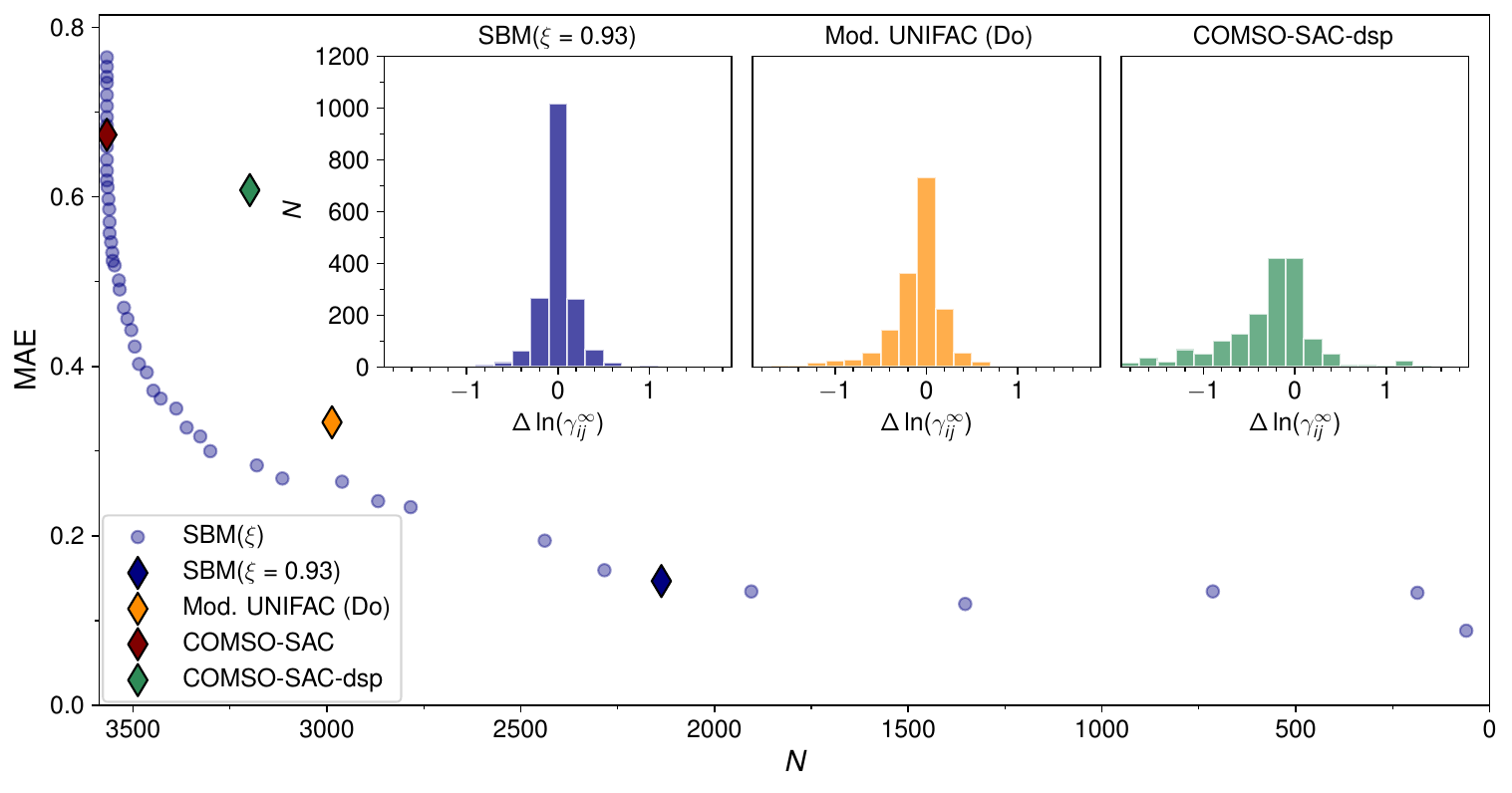}
    \caption{Mean absolute error (MAE) of the best-performing SBM (with varied thresholds $\xi$), modified~UNIFAC (Dortmund), COSMO-SAC, and COSMO-SAC-dsp for the prediction of $\ln \gamma^\infty_{ij}$ over the number of predictable experimental data points $N$. Insets provide histograms of the deviations of the predictions with the SBM with $\xi=0.93$, modified~UNIFAC (Dortmund), and COSMO-SAC-dsp from the experimental data, considering only mixtures that all three methods can describe. The shown interval in the histograms contains 99.9~\% (SBM), 96.7~\% (modified~UNIFAC (Dortmund)), and 96.9~\% (COSMO-SAC-dsp) of the relevant 1,748 data points.}
    \label{Sim1_fig:pareto_histograms}
\end{figure}

The best-performing SBM ($w_\sigma = 0.6$, and $w_\mathrm{P} = 2$) selected in the grid search is further evaluated in the following by comparison against the state-of-the-art physical benchmark methods for predicting activity coefficients: modified~UNIFAC (Dortmund)~\cite{Constantinescu.2016}, COSMO-SAC~\cite{Hsieh.2010}, and COSMO-SAC-dsp~\cite{Hsieh.2014}. As shown in Fig.~\ref{Sim1_fig:pareto_histograms}, the methods are compared using the MAE and the scope regarding the number of predictable data points $N$ in our data set. Additionally, the deviations of the predictions from the experimental data are plotted in histograms for the SBM with $\xi=0.93$, modified~UNIFAC (Dortmund), and COSMO-SAC-dsp. Modified~UNIFAC (Dortmund) has some extreme outliers, which were excluded from the MAE calculations in Fig.~\ref{Sim1_fig:pareto_histograms}. A detailed analysis of these outliers can be found in the Supporting Information, cf.~Table S.1.
In Fig.~\ref{Sim1_fig:pareto_histograms}, the scope is discussed regarding the number of predictable data points from our experimental database. An additional discussion of scope, in terms of the filling level of the entire matrix as shown in Fig.~\ref{Sim1_fig:Heatmap_Exp}, i.e., the predictions for mixtures for which no experimental data are available, is provided in the Supporting Information.

First of all, it is evident from Fig.~\ref{Sim1_fig:pareto_histograms} that for the physical models, there is also a trade-off between the scope of the method and its accuracy. COSMO-SAC-dsp is more accurate than COSMO-SAC, but in its current parameterization~\cite{Bell.2020}, it is not applicable to components containing certain halogens due to missing parameters for the dispersion part, resulting in a slightly smaller scope. Both COSMO variants have a larger scope than modified UNIFAC (Dortmund) but achieve less accurate results. 

Compared to each physical benchmark method, one can always find an SBM variant (by varying $\xi$) that outperforms it in terms of prediction accuracy and scope by selecting an appropriate threshold. Specifically, at $\xi=0.62$, the SBM can, like COSMO-SAC, predict all binary systems in our database but achieves a better MAE (0.62 compared to 0.67). At $\xi=0.85$, the SBM has a broader scope than COSMO-SAC-dsp ($N=3,301$ compared to $N=3,199$) and achieves a better MAE (0.30 compared to 0.61). Similarly, the SBM with $\xi=0.87$ has a broader scope than modified UNIFAC (Dortmund) ($N=3,115$ compared to $N=2,987$) and achieves a better MAE (0.27 compared to 0.33).

For the following analysis, we fix the threshold to $\xi=0.93$. While this value is, in principle, arbitrary, the resulting model can predict more than half of the available experimental data in our database with relatively high predictive accuracy. The deviations of the predictions from the experimental data for each method are also represented as histograms in Fig.~\ref{Sim1_fig:pareto_histograms}. Most of the predictions of the SBM with $\xi=0.93$ show deviations from experimental data smaller than $\pm$0.1, which is within the typical range of experimental uncertainty of $\ln \gamma^\infty_{ij}$, underscoring the high quality of the predictions that can be obtained with the proposed model.

\begin{figure}[H]
    \centering
    \includegraphics[width=\textwidth]{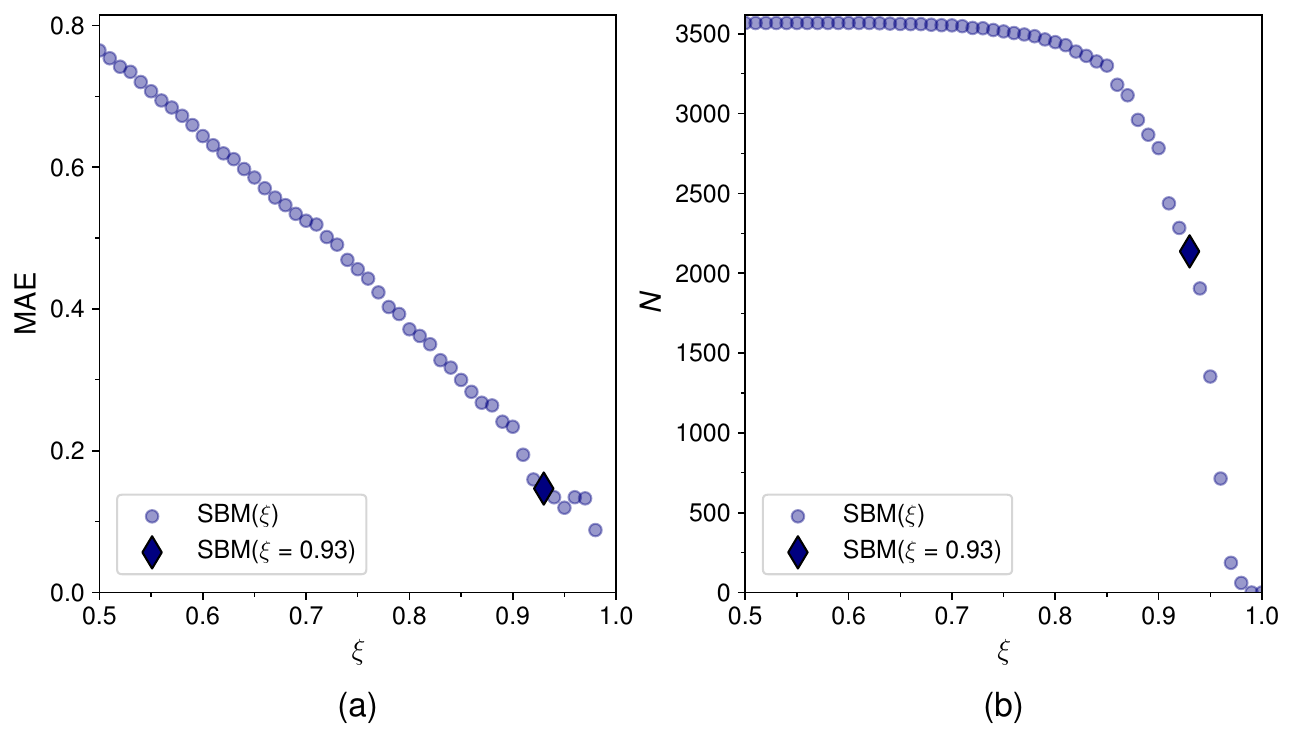}
    \caption{Mean absolute error (MAE) for the prediction of $\ln \gamma^\infty_{ij}$ (panel a) and number of predictable experimental data points $N$ (panel b) of the best-performing SBM over the threshold $\xi$. The results for the SBM with $\xi=0.93$ are highlighted.}
    \label{Sim1_fig:MAEandN_vs_xi}
\end{figure}

To further analyze the performance of the best-performing SBM, we plot the respective objectives (MAE and $N$) over the threshold $\xi$, as shown in Fig.~\ref{Sim1_fig:MAEandN_vs_xi}.
Fig.~\ref{Sim1_fig:MAEandN_vs_xi}a shows that increasing $\xi$ results in a nearly linear decrease in MAE, indicating improving accuracy. In contrast, the relationship between $N$ and $\xi$ in Fig.~\ref{Sim1_fig:MAEandN_vs_xi}b is more complex. For $\xi\leq0.62$, the model achieves its maximum scope, i.e., predicting all experimental data points, while for $\xi>0.98$, none of the mixtures can be predicted. Between these two boundaries, $N$ first decreases slowly with increasing $\xi$, followed by a steep decrease as $\xi$ approaches 1. This sensitivity of $N$ to $\xi$ emphasizes the importance of selecting an optimal threshold. Overall, Fig.~\ref{Sim1_fig:MAEandN_vs_xi} supports the choice of $\xi=0.93$, marked by the diamond, as a balance point that combines high predictive accuracy with substantial scope. While selecting a lower threshold would yield a broader scope, $\xi=0.93$ is preferred here as it achieves an MAE in the range of typical experimental uncertainties.

\begin{figure}[H]
    \centering
    \includegraphics[width=\textwidth]{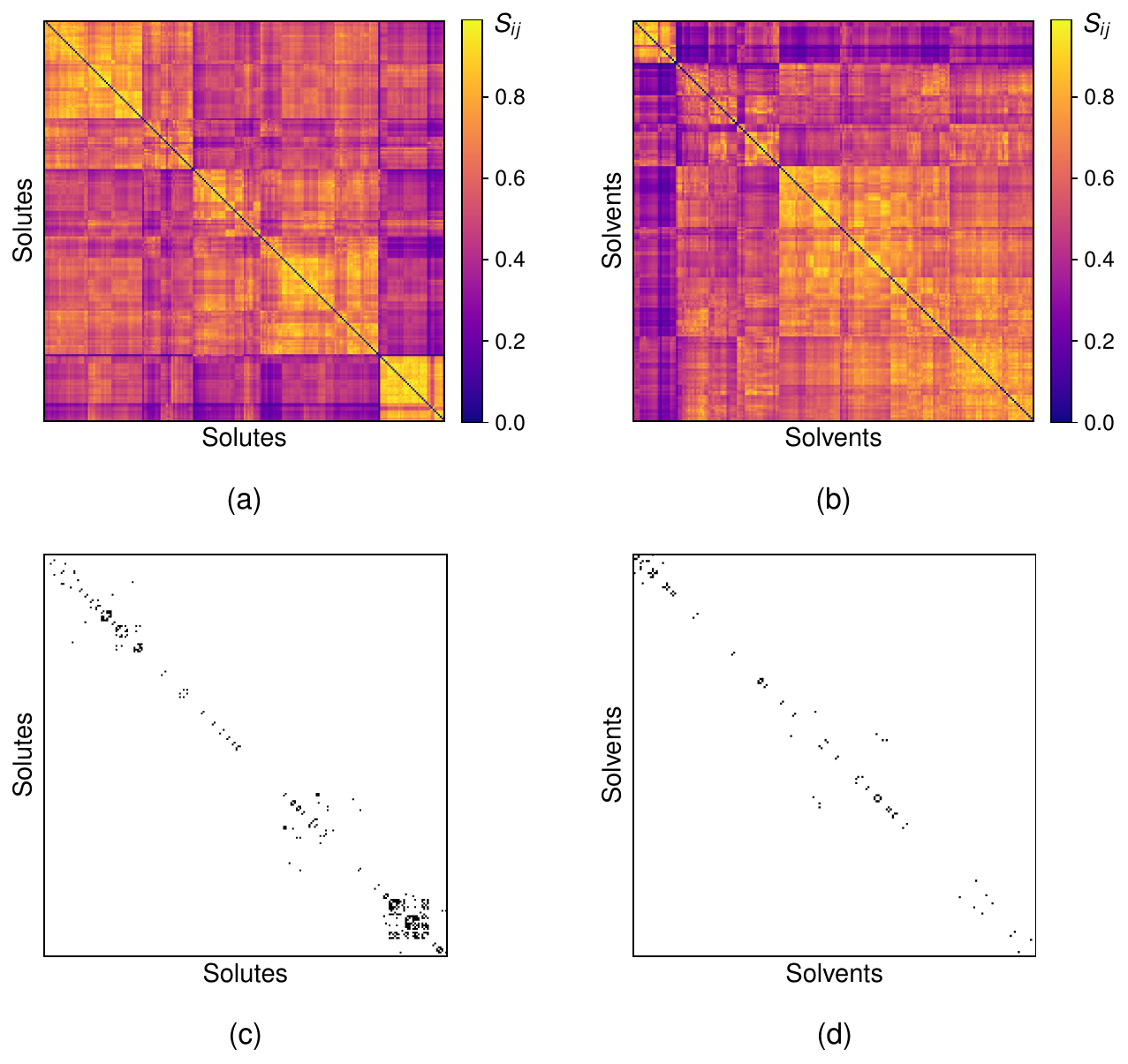}
    \caption{Heatmaps showing results for the pairwise similarity scores $S_{mn}$ of the considered solutes (panel a) and solvents (panel b). For illustration, pairs with $S_{mn} > 0.93$ are highlighted in panels c and d.}
    \label{Sim1_fig:SimMatrices}
\end{figure}

A detailed analysis of the results for the similarity $S_{mn}$ of all pairs of solutes and all pairs of solvents is presented in Fig.~\ref{Sim1_fig:SimMatrices}. The results are plotted in matrices, which are symmetric as $S_{mn}=S_{nm}$. In these matrices, the solutes (solvents) were arranged so that similar solutes (solvents) were positioned nearby, which was done using a clustering algorithm adopted from a previous work\cite{Gond.2024}. The chosen arrangement of the solutes (solvents) leads to high values of $S_{mn}$ along the diagonal, cf.~Fig.~\ref{Sim1_fig:SimMatrices}.
The heatmaps in Fig.~\ref{Sim1_fig:SimMatrices}a and b reveal only a few strong similarities among the solutes and solvents in our database, as indicated by the few bright yellow areas. A notable exception is observed in the lower right corner of the solute matrix, cf.~Fig.~\ref{Sim1_fig:SimMatrices}a, where a yellow square primarily represents alkanes classified as very similar according to our metric.

This observation becomes even more apparent when highlighting the solute-solute and solvent-solvent combinations with $S_{mn}$ higher than $\xi=0.93$, the threshold chosen for the detailed analysis discussed above, cf.~Fig.~\ref{Sim1_fig:SimMatrices}c and d. Interestingly, only very few, or even just one, similar solutes or solvents for the mixture of interest are needed for the SBM to achieve the excellent predictive accuracy discussed earlier. Thus, for a set of similar mixtures, i.e., those with at least one similar solute or solvent according to our similarity metric, it is sufficient to measure $\ln \gamma^\infty_{ij}$ for just one of them. The other mixtures can then be predicted with high accuracy using the SBM. This finding is exciting for the planning of experiments in several ways. For example, it opens up ways to replace substances that are difficult to handle experimentally by suitable proxies, and it can also be used to devise strategies for an efficient design of experiments (DOE) to improve the accuracy and scope of the SBM with a minimum amount of additional experimental data.

\clearpage
\section{Conclusions}
This work has two primary outcomes: the first is a new way to measure the similarity between two components. It only needs the components' $\sigma$-profiles and their surface areas as input, information that can be obtained for basically any molecule from a quantum-chemical calculation or databases. Hence, the new measure of similarity is highly versatile. We compare the information on these two properties of the two components $m$ and $n$ and combine the result in a similarity score $S_{mn}$, which is defined in such a way that the values always range between 0 and 1. The definition of this score contains hyperparameters (weights) that can be adapted to different tasks. In the present study, our goal was to use $S_{mn}$ to develop a new method for predicting the limiting activity coefficient $\gamma^\infty_{ij}$ of a solute $i$ in a solvent $j$ in systems for which no experimental data are available. We have chosen the hyperparameters so that the resulting similarity scores are beneficial for this task. However, the resulting definition of the similarity score should also be helpful for many other tasks related to predicting or assessing the thermodynamic properties of binary liquid systems.

The second outcome of the work is the new similarity-based method (SBM) for predicting isothermal activity coefficients at infinite dilution $\gamma^\infty_{ij}$. The idea behind the method is simple. We start with a database on $\gamma^\infty_{ij}$ at the temperature of interest and want to predict the value for a certain combination $i+j$ for which no data are available. Let us assume we have a data point for $\gamma^\infty_{in}$ for a system with the solute $i$ of interest in combination with another solvent $n$. We simply check whether the solvents $j$ and $n$ are sufficiently similar (i.e.,~$S_{jn} > \xi$) and then take the result for $\gamma^\infty_{in}$ as a proxy for $\gamma^\infty_{ij}$. (The same works if the problem is not the solvent but the solute). Of course, there must be rules on handling cases in which several such proxies are found. We apply a simple arithmetic average in this case, but taking the arithmetic average of the results of sufficiently similar substances is only one option; others could be explored. In the procedure of predicting $\gamma^\infty_{ij}$, another hyperparameter is introduced, the threshold $\xi$. We leave this threshold open, and the user can specify it. The functions that relate the chosen value of $\xi$ to the number of systems for which predictions are possible and the expected accuracy of the prediction (measured, e.g.,~by the MAE obtained in a leave-one-out study) can be easily established, and give guidance for the application of the method. In general, the higher $\xi$, the more accurate the prediction will be, but high values of $\xi$ will compromise the method's applicability. 

The SBM we have developed here for predicting $\gamma^\infty_{ij}$ shows a remarkable accuracy, even though the database is not large and typically contains only very few (if any) highly similar systems for any given combination of solute $i$ and solvent $j$. The new SBM outperforms the established physical benchmark methods UNIFAC (Dortmund), COSMO-SAC, and COSMO-SAC-dsp. 

The approach for designing SBMs based on our new similarity score $S_{mn}$ is generic and can be transferred to any physical property of binary liquid mixtures. For thermodynamic applications, the hyperparameters of $S_{mn}$ determined in the present work should be a good starting point but could be adapted for other applications.

The observation that data for only a few similar mixtures are sufficient to achieve accurate predictions suggests that a comparatively low number of targeted experiments can considerably improve SBMs. More generally, this finding could form the basis for new guiding principles for the design of experiments in binary systems.

\section{Data Availability Statement}
The experimental data on limiting activity coefficients were used under license for this study; they are available directly from Dortmund Data Bank (DDB) version 2023 \cite{DDB2023}. The $\sigma$-profiles as well as the implementations of COSMO-SAC and COSMO-SAC-dsp were taken from the open-source database provided by Bell et al.~\cite{Bell.2020}.
\section{Conflicts of Interest}
There are no conflicts of interest to declare.
\begin{acknowledgement}
The authors gratefully acknowledge financial support by Carl Zeiss Foundation in the frame of the project 'Process Engineering 4.0' and by DFG in the frame of the Priority Program SPP2363 'Molecular Machine Learning' (grant number 497201843). Furthermore, FJ gratefully acknowledges financial support by DFG in the frame of the Emmy-Noether program (grant number 528649696).
\end{acknowledgement}
\clearpage
\providecommand{\latin}[1]{#1}
\makeatletter
\providecommand{\doi}
  {\begingroup\let\do\@makeother\dospecials
  \catcode`\{=1 \catcode`\}=2 \doi@aux}
\providecommand{\doi@aux}[1]{\endgroup\texttt{#1}}
\makeatother
\providecommand*\mcitethebibliography{\thebibliography}
\csname @ifundefined\endcsname{endmcitethebibliography}  {\let\endmcitethebibliography\endthebibliography}{}

\end{document}


\section{Outliers of Modified UNIFAC (Dortmund)}
The predictions of modified UNIFAC (Do)\cite{Constantinescu.2016} include eight extreme outliers, cf.~Table~\ref{Sim1_tab:outliers_modUNIFAC}, which can be attributed to poorly fitted group-interaction parameters. Specifically, all of the relevant solutes contain main group 42 ("CY-CH2"), while all of the relevant solvents contain main group 18 ("PYRIDINE").

The few observed outliers would drastically increase the mean absolute error (MAE) and mean squared error (MSE) and thus lead to a false impression of the predictive performance of modified UNIFAC (Dortmund); they were therefore removed from this study. By removing the eight listed outliers, the MAE decreases from 0.6477 to 0.3340.
\begin{table}[H]
	\centering
	\caption{Binary systems with available experimental $\ln \gamma^\infty_{ij}$ at 298.15~K where modified UNIFAC (Dortmund) predictions deviate significantly from the experimental data, likely due to inaccurate group-interaction parameters. Identifiers (DDB no.) are the original ones from the DDB~\cite{DDB2023}.}
	\label{Sim1_tab:outliers_modUNIFAC}
	\begin{tabular}{c c c c}
		\toprule
		\multicolumn{2}{c}{Solute $i$} & \multicolumn{2}{c}{Solvent $j$} \\
		DDB no. & Name & DDB no. & Name \\
		\midrule
        50 & Cyclohexane & 19 & 2-Methylpyridine\\
        50 & Cyclohexane & 144 & Pyridine\\
        50 & Cyclohexane & 433 & Quinoline\\
        51 & Cyclopentane & 433 & Quinoline\\
        52 & Cyclohexene & 144 & Pyridine\\
        159 & Tetrahydrofuran & 144 & Pyridine\\
        401 & Ethylcyclohexane & 19 & 2-Methylpyridine\\
        401 & Ethylcyclohexane & 144 & Pyridine\\
		\bottomrule
	\end{tabular}
\end{table}

\clearpage
\section{Results of the Hyperparameter Variations}
\begin{figure}[H]
    \centering
    \includegraphics[width=0.5\textwidth]{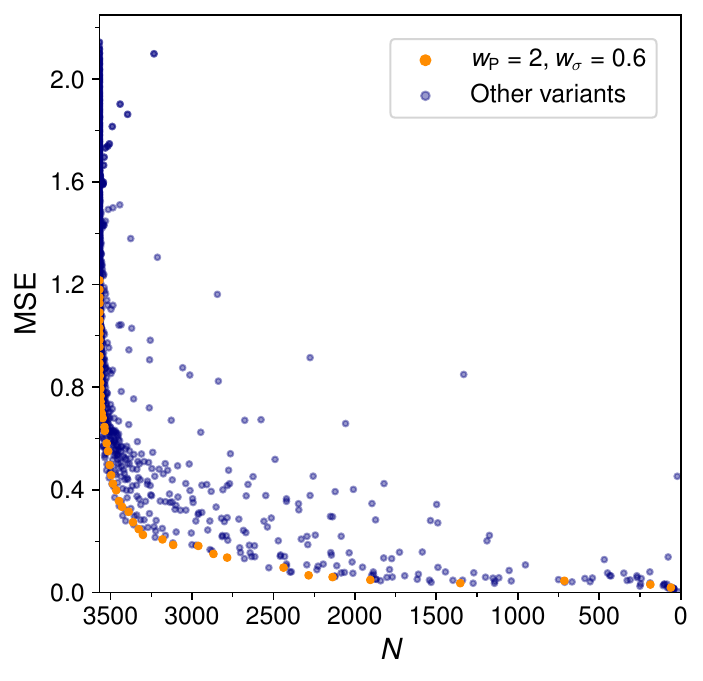}
    \caption{Mean squared error (MSE) of the predicted $\ln \gamma^\infty_{ij}$ over the number of predictable experimental data points $N$, as defined by the threshold $\xi$, for all tested SBM variants. The results of the best-performing SBM (as specified with the weights $w$) are highlighted in orange.}
    \label{Sim1_fig:GridSearch_MSE}
\end{figure}

The predictive performance of each SBM variant is shown in Fig.~3 in the manuscript, focusing on the MAE. By displaying the MSE in a similar way, Fig.~\ref{Sim1_fig:GridSearch_MSE} supports the choice of the final model, represented by the orange dots. 

\begin{figure}[H]
    \centering
    \includegraphics[width=\textwidth]{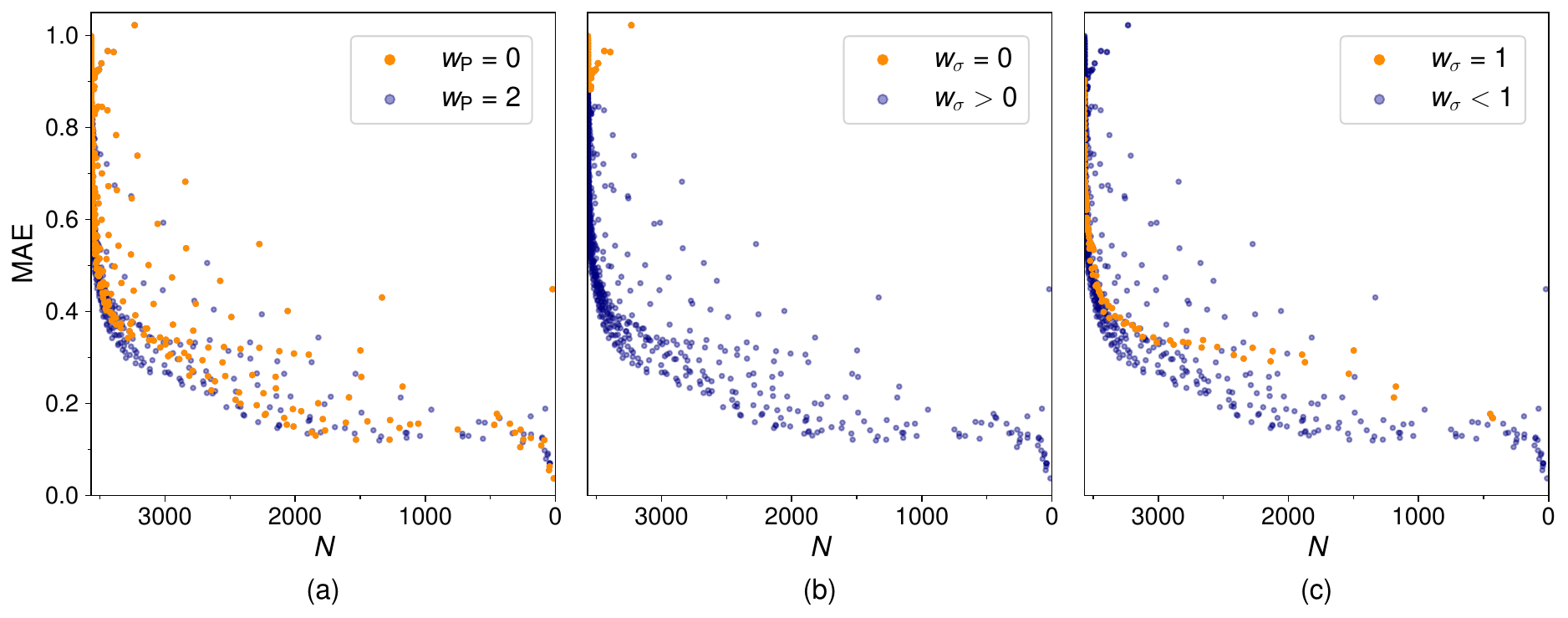}
    \caption{Mean absolute error (MAE) of the predicted $\ln \gamma^\infty_{ij}$ over the number of predictable experimental data points $N$, as defined by the threshold $\xi$, for all tested SBM variants. In each panel, a subset of SBM variants is highlighted in orange: (a) equal weighting of the polar and non-polar regions in the $\sigma$-profiles ($w_\mathrm{P}=0$), (b) using only the surface area similarity ($w_\sigma=0$), (c) using only the charge distribution similarity ($w_\sigma=1$).}
    \label{Sim1_fig:GridSearch_MAE_Heuristics}
\end{figure}

In Fig.~\ref{Sim1_fig:GridSearch_MAE_Heuristics}, a detailed analysis of the results is shown, focusing on the influence of different hyperparameter choices. From this analysis, we derive the following heuristics:
\begin{itemize}
    \item A stronger weighting of the polar regions in the $\sigma$-profiles ($w_\mathrm{P}=2$) enhances both objectives, i.e., the predictive performance and the scope, especially near the Pareto knee, cf.~Fig.~\ref{Sim1_fig:GridSearch_MAE_Heuristics}a.
    \item Relying solely on surface area similarity ($w_\sigma=0$) results in poor predictions, cf.~Fig.~\ref{Sim1_fig:GridSearch_MAE_Heuristics}b.
    \item Focusing solely on charge distribution similarity ($w_\sigma=1$) achieves comparatively good results at lower thresholds, where both the SBM scope and the MAE are generally large. However, increasing the threshold yields only small improvements in predictive accuracy, cf.~Fig.~\ref{Sim1_fig:GridSearch_MAE_Heuristics}c. In contrast, many model variants incorporating surface area similarity ($w_\sigma<1$) perform significantly better here.
\end{itemize}

\clearpage
\section{Scope of the Proposed Similarity-Based Method}
\begin{figure}[H]
    \centering
    \includegraphics[width=\textwidth]{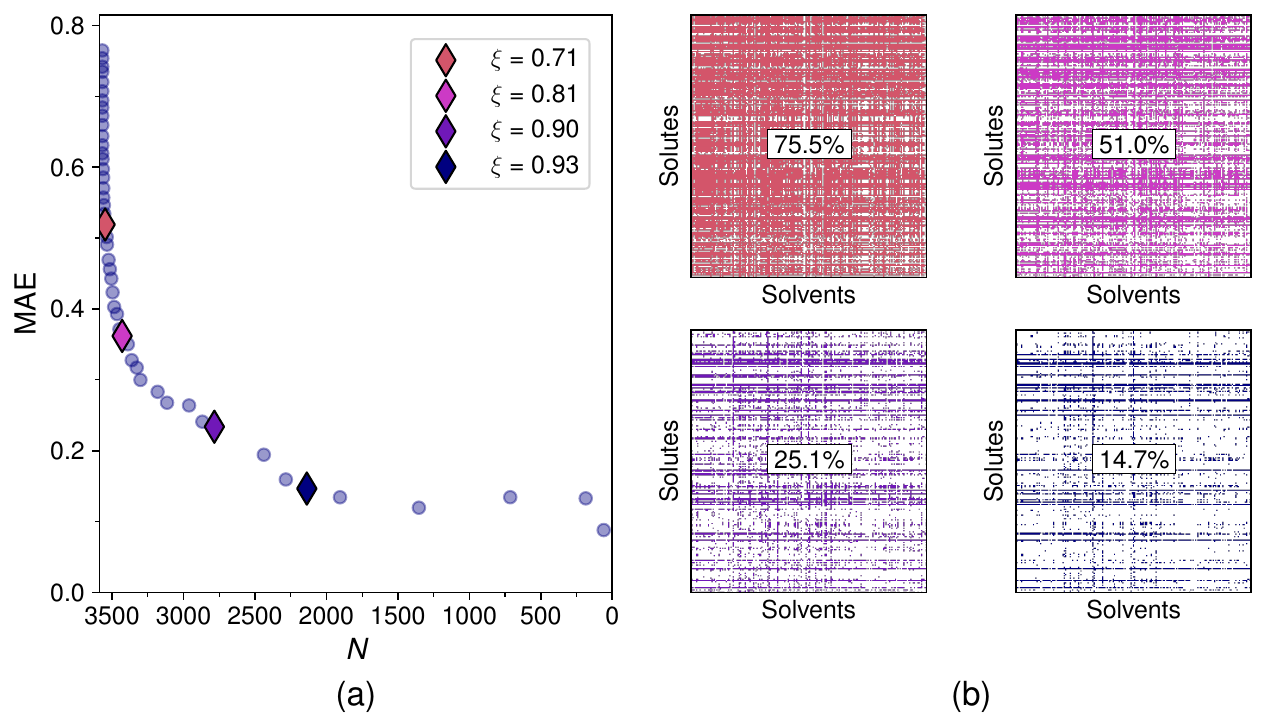}
    \caption{Influence of the threshold $\xi$ on the predictive performance and the scope of the SBM. (a) Mean absolute error (MAE) of the SBM for the prediction of $\ln \gamma^\infty_{ij}$. $N$ represents the number of predictable experimental data points. Four distinct thresholds are highlighted. (b) Matrices representing all predictable solute-solvent combinations for the four $\xi$ values highlighted in (a).}
    \label{Sim1_fig:pareto_matrices}
\end{figure}

The performance of the final SBM, characterized through $w_\sigma = 0.6$, and $w_\mathrm{P} = 2$, can be influenced by varying the threshold $\xi$. Fig.~\ref{Sim1_fig:pareto_matrices} illustrates the resulting trade-off between scope and predictive accuracy, focusing not only on the scope in terms of the number of predictable data points from the database (Fig.~\ref{Sim1_fig:pareto_matrices}a), but also on the number of predictable data points from all possible solute-solvent combinations (Fig.~\ref{Sim1_fig:pareto_matrices}b).
In general, a large $\xi$ results in a small scope and vice versa. Thereby, the percentage of predictable data points from the experimental database is always higher than the scope concerning the entire solute-solvent matrix. For example, the selected threshold of $\xi=0.93$ yields an SBM that can predict 59.9\% of the experimental database but can only populate 14.7\% of the solute-solvent matrix. The SBM is limited in extrapolating into the sparse region of the matrix, which is not surprising since it relies on experimental data points of similar mixtures. In comparison, modified UNIFAC (Do)~\cite{Constantinescu.2016} is capable of predicting 83.7\% of the experimentally studied mixtures and 69.9\% of the entire matrix, COSMO-SAC-dsp~\cite{Hsieh.2014} achieves 89.7\% and 85.4\%, respectively, whereas COSMO-SAC~\cite{Hsieh.2010} can calculate $\ln\gamma^\infty_{ij}$ for all considered binary mixtures.

By reducing $\xi$, the scope of the SBM could be extended so that almost any solute-solvent combination can be predicted. However, this would lead to a significant loss of predictive accuracy. Therefore, we continue to use the SBM with $\xi=0.93$ as its prediction accuracy is within the range of experimental uncertainty, and we accept the limited scope. The unprecedented performance of this SBM makes it a precious tool in chemical process engineering, even if it is not applicable in all cases.

\clearpage
\section{Case Studies of Similar Components}
The calculated similarity scores $S_{ij}$ between two components are not only the basis of the proposed SBM for the prediction of activity coefficients at infinite dilution. They also offer the possibility to identify the most similar components for a target component, exemplified in Tables \ref{Sim1_tab:similarSolutes} and \ref{Sim1_tab:similarComp_H2OandEtOH}. Here, the $S_{ij}$ of the final SBM, obtained through the grid search, are used.
\begin{table}[H]
	\centering
	\caption{Lists of top 10 components among the solutes most similar to ethanol or n-butane.}
	\label{Sim1_tab:similarSolutes}
	\begin{tabular}{c c p{1em} c c}
		\toprule
		\multicolumn{2}{c}{Solutes Similar to Ethanol} & & \multicolumn{2}{c}{Solutes Similar to n-Butane}\\
		Solute & $S_{ij}$ & & Solute & $S_{ij}$ \\
		\midrule
        1-Propanol & 0.909 & & Cyclopentane & 0.962\\
        2-Propanol & 0.869 & & 2-Methylpropane & 0.961\\
        Methanol & 0.850 & & Pentane & 0.930\\
        1-Butanol & 0.835 & & 2-Methylbutane & 0.929\\
        N-Methylformamide & 0.829 & & Propane & 0.917\\
        2-Butanol & 0.808 & & Methylcyclopentane & 0.910\\
        1-Pentanol & 0.806 & & Cyclohexane & 0.894\\
        tert-Butanol & 0.801 & & 3-Methylpentane & 0.890\\
        2-Methyl-1-propanol & 0.776 & & 2-Methylpentane & 0.886\\
        Cyclohexanol & 0.770 & & 2,3-Dimethylbutane & 0.881\\
		\bottomrule
	\end{tabular}
\end{table}

\begin{table}[H]
	\centering
	\caption{Lists of top 10 components among the solvents most similar to water or chlorobenzene.}
	\label{Sim1_tab:similarComp_H2OandEtOH}
	\begin{tabular}{c c p{1em} c c}
		\toprule
		\multicolumn{2}{c}{Solvents Similar to Water} & & \multicolumn{2}{c}{Solvents Similar to Chlorobenzene}\\
		Solvent & $S_{ij}$ & & Solvent & $S_{ij}$ \\
		\midrule
        Methanol & 0.716 & & Bromobenzene & 0.977\\
        Formamide & 0.702 & & Iodobenzene & 0.948\\
        Ethanol & 0.636 & & Fluorobenzene & 0.921\\
        1,2-Ethanediol & 0.623 & & 1-Chloronaphthalene & 0.869\\
        1-Propanol & 0.599 & & 1-Bromonaphthalene & 0.861\\
        2-Propanol & 0.592 & & 1,1-Dichloroethane & 0.770\\
        1,3-Propanediol & 0.591 & & Toluene & 0.746\\
        1-Butanol & 0.572 & & Methoxybenzene & 0.743\\
        1,4-Butanediol & 0.572 & & Indene & 0.742\\
        1,2-Propanediol & 0.569 & & Diiodomethane & 0.740\\
		\bottomrule
	\end{tabular}
\end{table}
Not surprisingly, many alkanes are similar to n-butane, and many halogenobenzenes are similar to chlorobenzene. In contrast, water has no similar components, as it is unique in being an extremely polar and rather small molecule.
\clearpage
\providecommand{\latin}[1]{#1}
\makeatletter
\providecommand{\doi}
  {\begingroup\let\do\@makeother\dospecials
  \catcode`\{=1 \catcode`\}=2 \doi@aux}
\providecommand{\doi@aux}[1]{\endgroup\texttt{#1}}
\makeatother
\providecommand*\mcitethebibliography{\thebibliography}
\csname @ifundefined\endcsname{endmcitethebibliography}  {\let\endmcitethebibliography\endthebibliography}{}